\documentclass[11pt,preprint]{aastex}
\usepackage{epsfig}
\begin{document}
\title{Spin-down of compact stars and energy release of a first-order phase transition}
\author{Kang Miao$^{1,2}$, Zheng Xiao-Ping$^{1}$, Pan
Na-Na$^{1}$\\
1 - {\it  The institute of astrophysics,Huazhong normal university},  Wuhan 430079, Hubei, P.R.China.\\
2 - {\it The college of physics and electron,Henan university},
Kaifeng, 475004, Henan, P.R.China.}

\begin{abstract}
  The deconfinement phase transition from hadronic matter to quark matter
can continuously occur during spins down of neutron stars. It will lead to the release
 of latent heat if the transition is the first-order one. We have investigated the energy release
of such deconfinement phase transition for rotating hybrid stars
model which include mixed phase of hadronic matter and quark
matter.
 The release of latent heat per baryon is calculated through studying a
 randomly process of infinitesimal compressing.
Finally,  we can self-consistently get the heating luminosity of
 deconfinement phase transition by imputing the
 EOS of mixed phase, and based on the equation of rotation structure of stars.

\bigskip
\end{abstract}
\section*{1. Introduction}

\parskip 2pt
Neutron stars (NSs) provide us a unique playground to study the
properties of super-dense matter in the most extreme physical
conditions. NSs participation in various astrophysical phenomena
usually presented challenges for us. The different equation of state
(EOS) and their properties of matter at high densities would play
an important role in understanding some astrophysical phenomena.

One of the most intriguing predictions of some theories of dense
matter is a possibility of a phase transition into an 'exotic'
state, including pion and kaon condensation, and deconfinement of
quarks. Of them, the deconfinement phase transition represented
the most profound effect on the structure and dynamics of NSs.
Many investigations were interesting in the phase transition which
is of the first-order type(Pisalski $\&$ Wilczek \cite{84pis} and
Gavai et al. \cite{87gav}). In the simplest case, equilibrium
phase transition from the normal, lower density phase to the pure
exotic one, occurs at a constant pressure, and is accompanied by a
density jump at the phase interface(Baym \& Chin \cite{76bay}).
However, as shown by Glendenning (Glendenning
\cite{92gle},\cite{97gle}), the properties of phase transition
with more than one conserved charge are quite different from the
constant pressure transition. In such a case, a mixed phase(MP)
can be made in the interior NSs. The NSs containing MP matter is
so-called hybrid stars (HSs) by many investigators.

 It is known that a NS will spin down due to braking (e.g. electric-magnetic radiation or gravitational wave radiation).
 Deconfinement  transition can proceed to occur in HSs during the spin down. Such evolutionary processes induce not only
 the changes in stellar structure but also continuous release of latent heat. In the present paper, we calculate the energy
 release due to a phase transition in a rotating HS. The heat arises only at the phase interface for a constant pressure
 transition. In our case, the heat release will be distributed over a MP region. Our calculations are based on the perturbation
 theory developed by Hartle(Hartle \cite{67har}). Regard EOS for hadronic matter and strange quark matter as
 the input, we will self-consistently acquire the deconfinement heat luminosity with respect to the characteristic age of the
 star.

In this work, we
 choose the simplest possible nuclear matter composition, namely
neutrons, protons, electrons, and muons (npe$\mu$ matter) and
ignore superfluidity and superconductivity.

The paper is organized as follows.
 In Sec.2, we introduce notation and describe general properties
of deconfinement phase transition in stellar with particular
emphasis on the existence of the MP and varying pressure in
the MP. The rotating structure of star with deconfinement
phase transition are presented in Sec.3. In Sec.4 we present the
calculation of deconfinement phase transition heat luminosity
associated with EOS of the MP and structure of rotating star.
 The conclusion and discussions are summarized
in Sec. 5.

\section*{2.Deconfinement phase transition}
Quark deconfinement phase transition is expected to occur in
neutron matter at densities above the nuclear saturation density
$\rho_{B}=0.16 fm^{-3}$. Early works on the possible occurrence of
quark matter in neutron stars(Baym \& Chin \cite{76bay}) were
based on the assumption that hadron matter and quark matter were
both charge neutral(with only one independent chemical potential).
As a consequence, the transition was described using Maxwell
construction and the resulting picture of the star consisted of
quark matter core surrounded by a mantle of hadron matter, the two
phases being separated by a sharp interface.

 In the 90s, Glendenning(Glendenning \cite{92gle}, \cite{97gle}) pointed
out that this assumption is too restrictive. More generally, the
transition can
 through the formation of a MP of hadron matter and quark
matter, total charge neutrality being achieved by a positively
charged amount of hadron matter and a negatively charged amount of
quark matter.
 Therefore, at present, most of the approaches to deconfinement
matter in NS matter use a standard two-phase description
of EOS where the hadron phase(HP) and the quark phase(QP) are modelled separately and
resulting EOS of the MP is obtained by imposing Gibbs conditions
for phase equilibrium with the constraint that baryon number as
well as electric charge of the system are conserved (Glendenning
\cite{97gle}, schertler et al\cite{00sch}).

We have to deal with two independent chemical potentials
($\mu_{n},\mu_{e}$) if we impose the condition of weak
equilibrium. The Gibbs condition for mechanical and chemical
equilibrium at zero temperature between the HP and the QP reads
\begin{equation}
p_{HP}(\mu_{n},\mu_{e})=p_{QP}(\mu_{n},\mu_{e}).
\end{equation}
where $p_{HP}$ is the pressure of HP and $p_{QP}$ is the pressure
of QP. We use the EOS of  the relativistic mean field model
(Glendenning \cite{97gle}) for hadron matter and employ an
effective mass bag-model EOS for quark matter(Schertle et al.
\cite{97sch}). Only two independent chemical potentials remain
according to the corresponding two conserved charges of the
$\beta$-equilibrium system. The volume fraction occupied by QP for
every point on the MP curve
\begin{equation}
\chi=\frac{V_{QP}}{V_{HP}+V_{QP}}
\end{equation}
can be obtained by imposing the condition of global charge
neutrality in the MP
\begin{equation}
\chi q_{QP}+(1-\chi) q_{HP}=0.
\end{equation}
 Where $q$ denotes the charge density.
Finally, the total energy density $\epsilon$ and baryon number $\rho_{B}$
 can be calculated using
 \begin{equation}
\epsilon=\chi\epsilon_{QP}+(1-\chi)\epsilon_{HP}
\end{equation}
\begin{equation}
\rho_{B}=\chi \rho_{QP}+(1-\chi) \rho_{HP}.
\end{equation}
 Taking the
charge neutral EOS of the HP, Eq.(1), (2) and (3) for MP and the
charge neutral EOS of the QP, we can construct the full hybrid
star EOS. The resulting of chemical potentials are shown in Fig.1.
Below the charge neutral HP curve and above the charge neutral QP
curve the HP is positively charged ($q_{HP}>0$) and the QP is
negatively charged($q_{HP}<0$). Therefore, the charge of hadronic
matter can be neutralized in the MP by an appropriate amount of
quark matter. In the MP the volume proportion of quark phase is
monotonically increasing from $\mu_{n}^{[1]}$ to $\mu_{n}^{[2]}$. In Fig.2 we show
the model EOS with deconfinement transition which is the typical
scheme of a first order transition at finite density with MP. The phase transition construction
in a two-component system leads to continuously increasing pressure of the MP with increasing density.
We choose the parameters for hadronic matter EOS which have given by
Glendenning \cite{97gle} and quark matter EOS with s quark mass
$m_{s}=150$MeV, bag constant $B^{1/4}=160$MeV, coupling constant
$g=3$.

\section*{3.Rotating evolution of hybrid stars}

With the evaluated hybrid star EOS presented above we now turn to
analyse the structure of the corresponding rotating HSs. Using the
Hartle's perturbation theory \cite{67har}, Chubarian et al
\cite{00chu} have studied the change of the internal structure of
the HSs due to rotation.
 In this paper, we
also apply Hartle's approach to investigate the structure of
rotating HSs. Hartle's formalism is based on treating a
rotating star as a perturbation on a non-rotating star, expanding
the metric of an axially symmetric rotating star in even powers of
the angular velocity $\Omega$. The metric of a slowly rotating
star to second order in the angular velocity $\Omega$, can be
written as
\begin{eqnarray}
ds^{2}=-e^{\nu(r)}[1+2(h_{0}+h_{2}P_{2})]dt^{2}+
e^{\lambda(r)}[1+\frac{2(m_{0}+m_{2}P_{2})}{(r-2M(r))}]dr^{2} \nonumber\\
+r^{2}[1+2(v_{2}-h_{2})P_{2}]
\{d\theta^{2}+\sin^{2}\theta[d\phi-w(r,\theta)dt]^{2}\}+O(\Omega^{3})
\end{eqnarray}
Here $e^{\nu(r)}$, $e^{\lambda(r)}$ and $M(r)$ are functions of
$r$ and describe the non-rotating star solution of the
Tolman-Oppenheimer-Volkov (TOV) equations(Oppenheimer\& Volkoff \cite{39opp}). $P_{2}=P_{2}(\theta)$
is the $l=2$  Legendre polynomials. $\omega$ is the angular
velocity of the local inertial frame and is proportional to the
star's angular velocity $\Omega$, whereas the perturbation
functions $h_{0},h_{2},m_{0},m_{2},v_{2}$ are proportional to
$\Omega^{2}$. we assume that matter in the star is described by a
perfect fluid with energy momentum tensor
\begin{equation}
T^{\mu\nu}=(\epsilon+P)u^{\mu}u^{\nu}+Pg^{\mu\nu}
\end{equation}
The energy density and pressure of the fluid are affected by the
rotation because the rotation deforms the star. In the interior of
the star at given $(r,\theta)$, in a reference frame that is
momentarily moving with the fluid, the pressure and energy density
variation is respectively
\begin{equation}
\delta P(r,\theta)=[\epsilon(r)+P(r)][ p_{0}^{*}+
p_{2}^{*}P_{2}(\theta)]
\end{equation}
\begin{equation}
\delta
\epsilon(r,\theta)=\frac{d\epsilon}{dP}[\epsilon(r)+P(r)][p_{0}^{*}+
p_{2}^{*}P_{2}(\theta)]
\end{equation}
here, $p_{0}^{*}$ and $p_{2}^{*}$ are dimensionless functions of
$r$, proportional to $\Omega^{2}$, which describe the pressure
perturbation. The rotational perturbations of the star's structure
are described by the functions
$h_{0},m_{0},p_{0}^{*},h_{2},m_{2},v_{2},p_{2}^{*}$. These
functions are calculated from Einstein's field equations. The
effect of rotation described by the metric on the shape of the
star can be divided into contributions: A spherical expansion
which changes the radius of the star, and is described by the
functions $h_{0}$ and $m_{0}$. The other part is a quadrupole
deformation, described by functions $h_{2}$, $v_{2}$ and $m_{2}$.
As a consequence of these contributions, the difference between
the gravitational mass of the rotating star and the non-rotating
star with the same central pressure is
\begin{equation}
\delta M_{grav}=m_{0}(R)+\frac{J^{2}}{R^{3}}
\end{equation}
The change in the radius of the star is given by
\begin{equation}
\delta R=\xi_{0}(R)+\xi_{2}(R)P_{2}(\theta)
\end{equation}
We wish to study sequences of the rotating stars with constant
total baryon number at variable spin frequency $\nu=\Omega/2\pi$.
The expansion of total baryon numbers in powers of $\Omega$ is
\begin{equation}
N_{B}=N_{B}^{0}+\delta N_{B}+O(\Omega^{4})
\end{equation}
where
\begin{equation}
N_{B}^{0}=\int_{0}^{R}n_{B}(r)[1-2M(r)/r]^{-1/2}4\pi r^{2}dr
\end{equation}
is the total baryons number of non-rotating star and
\begin{eqnarray}
\delta
N_{B}=\frac{1}{m_{N}}\int_{0}^{R}(1-\frac{2M(r)}{r})^{-1/2}\{[1+\frac{m_{0}(r)}
{r-2M(r)}+\frac{1}{3}r^{2}[\Omega-\omega(r)]^{2}e^{-\nu}]m_{N}n_{B}(r)\nonumber\\
+\frac{dm_{N}n_{B}(r)}{d P}(\epsilon+P) p_{0}^{*}(r)\}4\pi r^{2}dr
\end{eqnarray}
here $m_{N}$ is the rest mass per baryon. To construct constant
baryon number sequences, we first solve the TOV equations to find
the non-rotating configuration for giving a central pressure
P(r=0). And then, for an assigned value of the angular velocity
$\Omega$ the equations of star structure are solved to order
$\Omega^{2}$, imposing that the correction to the pressure $
p_{0}^{*}(r=0)$ being not equal to zero. The value of
$p_{0}^{*}(r=0)$ is then changed until the same baryon number as
that non-rotating star is obtained.

The results for the stability of rotating HSs
configurations with possible deconfinement phase transition
according to the EOS described above are shown in Fig.3, where the
total gravitation mass is given as functions of the equatorial
radius and the central baryon number density for static stars as
well as for stars rotating with the maximum rotation frequency
$\nu_{k}$. The dotted lines connect configuration with the same
total baryon number and it becomes apparent that the rotating
configurations are less compact than the static ones. In order to
explore the increase in central density due to spin down, we
create sequences of HSs models. Model in a particular
sequence have the same constant baryon number, increasing central
density and decreasing angular velocity. Fig.4 displays the
central density of rotating HSs with different
gravitational mass at zero spin, as a function of its rotational
frequency. In the interior of these stars, the matter can be
gradually converted from the relatively incompressible nuclear
matter phase to more compressible quark matter phase.

\section*{4. Deconfinement heating rate}
As the star spins-down, the centrifugal force decreases
continuously, increasing its internal density. Fig.4 identifies
the fact that quarks are accumulating in the interior of the star
with decreasing rotation frequency $\nu$. Since the deconfined
phase transition is first order phase transition, there are latent
heat produced with the transformation of hadron matter into quark
matter. The deconfinement phase transition heating play an
important role in the process of compact star's thermal evolution.

Corresponding to the EOS depicted in Fig.2, Fig.5 shows that
energy per baryon of HS matter as a function of the baryon number
density. Two intersectant solid lines denote the pure hadronic
matter phase and pure quark matter phase respectively. The
deconfinement transition just occurs at point 1 until point 2
after which a pure quark matter phase appears. Comparing HP curve
to MP one, we find that the enthalpy increase in MP is slower than
HP when baryon number density increases. It is a vivid
representation of latent heat release when a phase transition
occurs.

How to describe the latent heat is now a key issue. It is quite evident matter that the energy release due to the
phase transition is direct proportion to the difference between the HP derivative and the MP one at point 1. Similarly,
we can express the energy release per baryon as
\begin{equation}
\delta \tilde{e}-\delta e=(\frac{\delta
\tilde{e}}{\delta\rho_{B}}-\frac{\delta e}{\delta\rho_{B}})\delta
\rho_{B}
\end{equation}
where $\frac{\delta\tilde{e}}{\delta\rho_{B}}$ denotes the
enthalpy change per baryon for any density in MP curve, density
increase assumed no phase transition proceeds to occur. For
example, point A in Fig.5 has two possible enthalpy change ways,
AC and AD, when the density increases. No transition occurs along
AC while AD corresponds to the real situation. In order to derive
$\frac{\delta\tilde{e}}{\delta\rho_{B}}$,
 We rewrite
the expression of HP and QP volume
$V_{QP}=\frac{N_{QP}}{\rho_{QP}}$,$V_{HP}=\frac{N_{HP}}{\rho_{HP}}$
and define parameter $\eta=N_{Q}/N_{B}$. Substitute these equation
 into Eq.(2), we can get
\begin{equation}
\chi=\frac{\eta \rho_{HP}}{\eta \rho_{HP}+(1-\eta)\rho_{QP}}
\end{equation}
\begin{equation}
1-\chi=\frac{(1-\eta )\rho_{QP}}{\eta \rho_{HP}+(1-\eta)\rho_{QP}}
\end{equation}
 Replacing
Eq.(16) and Eq.(17) in Eq.(4) and Eq.(5), we have energy density for a given $\eta$,
\begin{equation}
\tilde{\epsilon}=\frac{\eta \rho_{HP}}{\eta
\rho_{HP}+(1-\eta)\rho_{QP}}\epsilon_{QP}+\frac{(1-\eta)
\rho_{QP}}{\eta \rho_{HP}+(1-\eta)\rho_{QP}}\epsilon_{HP}
\end{equation}
\begin{equation}
\tilde{\rho}=\frac{\rho_{QP}\rho_{HP}}{\eta
\rho_{HP}+(1-\eta)\rho_{QP}}
\end{equation}
In such a case, We can obtain the energy per baryon as
\begin{equation}
\tilde{e}=\frac{\tilde{\epsilon}}{\tilde{\rho}}=\eta
e_{QP}+(1-\eta)e_{HP}
\end{equation}
Furthermore we get
\begin{equation}
\frac{\partial\tilde{e}}{\partial \rho_{B}}=\eta\frac{\partial
e_{QP}}{\partial \rho_{B}}+(1-\eta)\frac{\partial e_{HP}}{\partial
\rho_{B}}
\end{equation}
Energy release  per baryon at point A in the process of
deconfinement phase transition can be given using above Eq.(15) and (21).
\begin{equation}
\Delta e=\delta\tilde{e}-\delta e=(\eta\frac{\partial
e_{QP}}{\partial \rho_{B}}+(1-\eta)\frac{\partial e_{HP}}{\partial
\rho_{B}}-\frac{\partial e}{\partial \rho_{B}})\delta \rho_{B}
\end{equation}

In Eq. (22), we see $\Delta e$ equals 0 when $\eta$ takes 1 or 0,
returning to pure phases, HP or QP. Integrating for whole star, we
get the total latent heat release unit time for a star
\begin{equation}
H=\frac{d\varepsilon}{dt}=\int \frac{de}{dt}\rho_{B}dV
\end{equation}
Since  the rotation frequency and its derivative can be observed, we can rewrite the Eq. (23) as
\begin{equation}
H=\int \frac{de}{d\nu}\dot{\nu}(t)\rho_{B}dV
\end{equation}

The heat release per baryon $\frac{de}{d\nu}$ due to reduction in
rotating frequency is simulated for a given rotating sequence
$N=1.66 N_{\odot}$ (the static mass $M=1.5 M_{\odot}$), is plotted
in Fig.6. We find that the energy release effectively is enhanced
with increasing density and frequency. So the deconfinement
transition in the star is stronger during the early ages, and the
energy release rate becomes larger and larger from the outer
region to the center of the star.

 Of course, the heat luminosity H can be obtained only if the spin-down rate of a star given. The common case is
 one induced by magnetic dipole radiation. It reads
\begin{equation}
\dot{\nu}=-\frac{16\pi^{2}}{3Ic^{3}}\mu^{2}\nu^{3}\sin^{2}\theta
\end{equation}
is induced by magnetic dipole radiation, where $I$ is the stellar
moment of inertia, $\mu=\frac{1}{2}BR^{3}$ is the magnetic dipole
moment, and $\theta$ is the inclination angle between magnetic and
rotational axes.

The total heat luminosity as a function of the characteristic age for different
 magnetic fields, for the rotating sequence $N=1.66 N_{\odot}$ (the static mass $M=1.5 M_{\odot}$), is presented in Fig.7.
 Heating source inside the stars is an important factor effect on the cooling of NSs. NSs first cools (for t $<10^{6}yrs$)
via various neutrino emission before the surface photon radiation take over.
Several heating mechanisms, for example, rotochemical
heating(Reisenegger et al. \cite{95rei},\cite{06rei}),
compositional transitions in crust (Iida \& Sato \cite{97iid}),
crust cracking (Cheng et al. \cite{92che}) and vortex pinning (Van
Riper et al. \cite{95van}), have been discussed in detail. It is
generaly expected that the heating sources significantly
contribute to surface emission for the old NSs with low magnetic
fields, characteristic of millisecond pulsars. Recently the
thermal emission data showed to have high surface temperature for
a few of millisecond pulsars. Especially for PSR J0437-4715, the
temperature can be inferred as $1.2\times 10^{5}K$. In light of
model for NS thermal evolution including rotochemical heating
source, Reisenegger (Reisenegger et al. \cite{06rei}) found that
the theoretical result is $20\%$ lower than the inferred
temperature. It is noteworthy in Fig.7 that our heat luminosity
for magnetic field $\sim10^{8}G$ and $\sim10^{9}G$ lasts during a
term far longer than $10^{6}yrs$ and is much higher than the other
heat generation. We thus think that high temperature of some
millisecond pulsars with low magnetic fields (Kargaltsev et al.
\cite{04kar}) can be explained using predicted heating model of
HSs. For PSR J0437-4715, we predict the surface emission
$L_{bol}=H(\nu,\dot{\nu})(1-\frac{2GM}{Rc^{2}})$ with rotation
period and its derivative $P=5.74,\dot{P}=3.64\times10^{-12}$. The
result is estimated as $L_{bol}\sim 6.6\times10^{29}erg s^{-1}$
considerably consistent with the observed thermal X-ray luminosity
$L_{X}=4\pi\sigma R^{2}T_{\infty}^{4}\sim 2.5\times 10^{29}erg
s^{-1}$.



\section*{5. Conclusions and discussions}
The nuclear matter can continuously be deconfined to quark matter
during the spins-down of star. The deconfinement phase transition
heating of rotating hybrid stars have been investigated in this
work. The latent heat release of such a first order phase
transition associated with the change of energy per baryon during
transition and rotational structure evolution of star. Using
Hartle's perturbative approach, we have calculated the change of
internal structure of rotating hybrid stars.  For a process of
infinitesimal compressing during the rotation of star, we can get
the energy release per baryon by inputing the EOS of the MP.
Furthermore, we
 self-consistently get the total energy release combining with the rotational structure of star.
The results show that the latent heat release per baryon per unit frequency enhances with
the increase of baryon number density and rotational frequency in HSs.

We have calculated the changes in the internal structure of a
compact star during its spin-down. As shown in previous studies
the deconfinement phase transition occurs if quark matter is
stable state at high densities, since the nuclear matter is
compressed in the interior of the star during spin-down. We
consider a first-order transition presented by Glendenning (1992)
to calculate the release of latent heat when the star is slow
down. The heating rate arising from the processes has been
estimated. We find its significant effects on the thermal
evolution and the effects is much more important than the past
present heating mechanisms. For old neutron stars with low
magnetic fields of $\sim 10^{8}-10^{9}G$, especially, our
predicted temperature is nearly identified with the values
observed from millisecond pulsars. In future, we expect more
observational examples in investigating effects of deconfinement
heating mechanism on the NSs thermal evolution. Thermal evolution
curves  need to be compared to more data. In this paper, for
simplicity we neglect Coulomb and surface effects on the MP. Many
investigations(Endo et al \cite{06end})have identified that both
effects would restrict the region of the MP in the core of hybrid
stars, which can effect the release of latent heat. Another
problem which remains to be investigated is the unified
description of middle-age and old pulsars. HS model may be no bad
selection when our combining deconfinement heating, superfluidity
effects in nuclear matter together.

\begin{acknowledgements}
This work is supported by NFSC under Grant Nos.10603002 and
10373007.
\end{acknowledgements}

%
%

  \begin{figure}
    \centering
    \includegraphics[height=80mm]{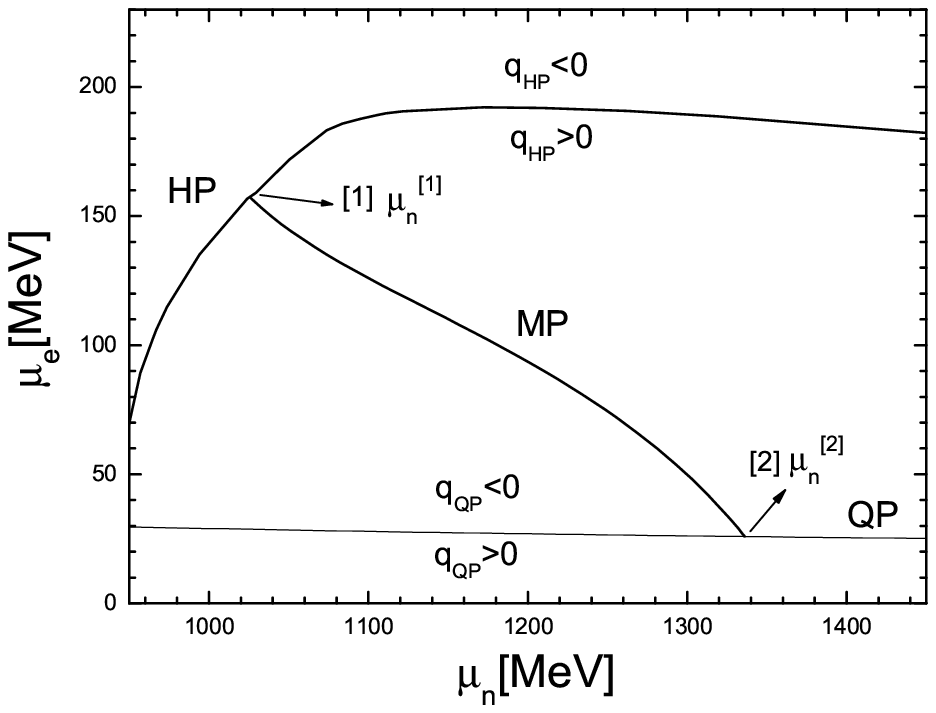}
    \caption{Electron chemical potential $\mu_{e}$ as a function of the neutron chemical potential $\mu_{n}$.
 The HP EOS is a
 relativistic mean-field model, the quark matter is
  effective mass MIT bag model with $m_{s}=150MeV$, $B^{1/4}=160 MeV$,coupling constant $g=3.0$.}

    \label{Fig:f1}
    \end{figure}

\begin{figure}
   \centering
   \includegraphics[height=80mm]{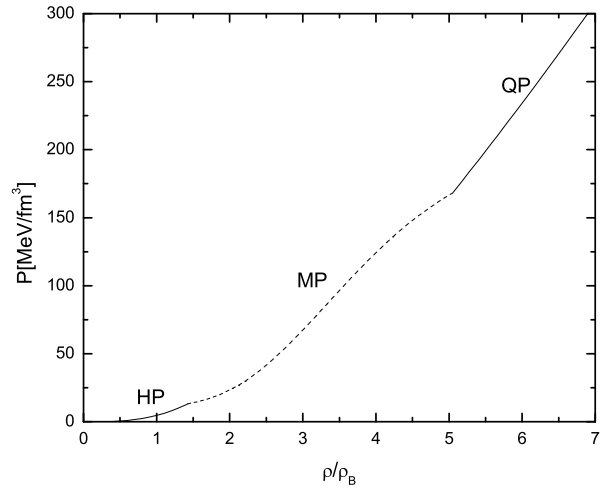}
   \caption{Model
EOS for the pressure of hybrid star matter
as a function of the baryon number density. The HP EOS is a
relativistic mean-field model, the quark matter is
 effective mass MIT bag model with $m_{s}=150MeV$, $B^{1/4}=160 MeV$,coupling constant $g=3.0$.}

   \label{Fig:f2}
   \end{figure}

 \begin{figure}
   \centering
   \includegraphics[height=80mm]{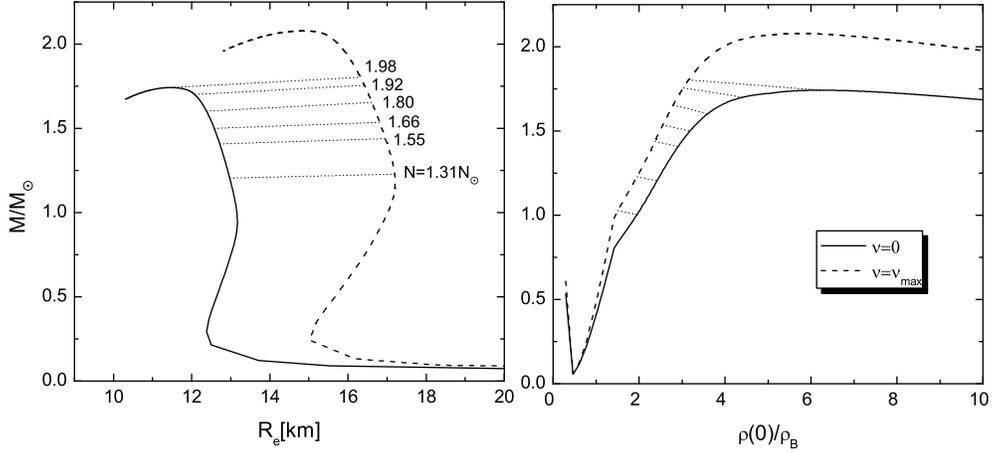}
   \caption{Gravitational mass M as a function of the equatorial radius (left figure) and the central
   density (right figure) for rotating hybrid stars configurations with a deconfinement phase phase transition.
 The solid curves correspond to static configurations. the dashed ones to those with maximum rotation frequency $\nu_{k}$.
 The lines between both extremal cases connect configurations with the same total baryon number.}
   \label{Fig:f3}
   \end{figure}

\begin{figure}
   \centering
   \includegraphics[height=80mm]{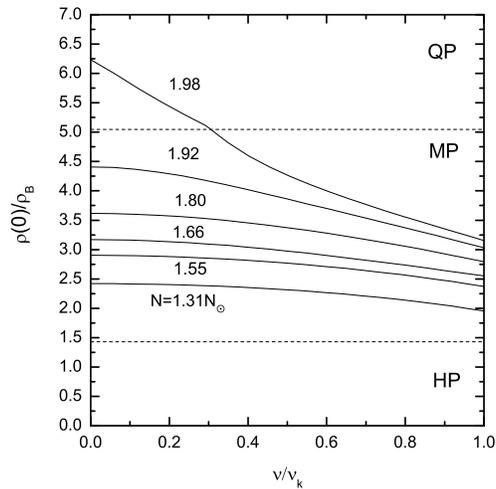}
   \caption{Central density as a function of rotational frequency for rotating hybrid
stars of different gravitational mass at zero spin. All sequences
are with constant total baryon number. Dash horizontal lines
indicate the density where quark matter is produced.
 }
   \label{Fig:f4}
   \end{figure}

\begin{figure}
\centering
   \includegraphics[height=80mm]{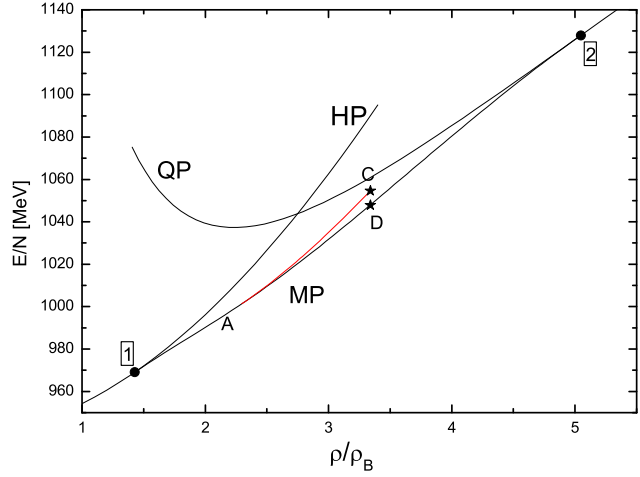}
   \caption{The energy of per baryon as a function of  the baryon number density
 for hybrid star matter. Two intersectant solid lines denote the pure
 hadronic matter phase and pure quark matter phase respectively. Line AC represents
the state of no deconfined phase transition during rotation of star.}
\label{Fig:f5}
\end{figure}

  \begin{figure}[htbp]
 \centering
     \includegraphics[height=80mm]{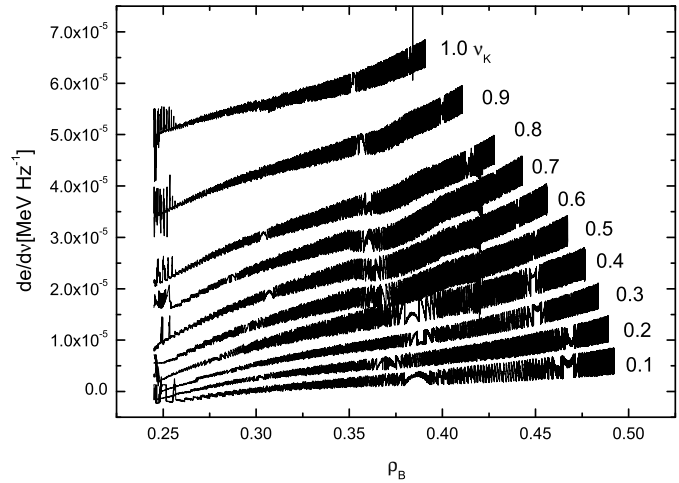}
    \caption{For 1.5 $M_{\odot}$ hybrid star,
the energy release of per baryon unit frequency as a function of
the baryon number density
  in mixed phase region for various frequency.}
  \end{figure}

 \begin{figure}
 \centering
    \includegraphics[height=80mm]{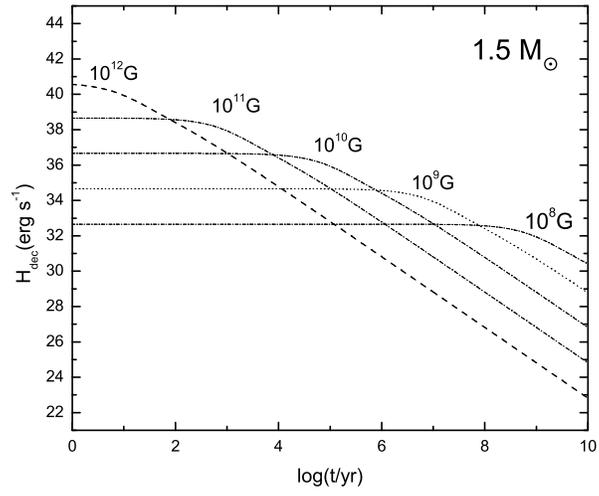}
    \caption{For 1.5 $M_{\odot}$ hybrid star, the deconfinement phase transition heating rate
  change with rotational evolution of star for various magnetic fields.}
    \label{Fig:f7}
    \end{figure}

%
\end{document}